\documentclass[a4paper]{article}
\begin{document}
\title{Extinction by grey dust in the intergalactic medium}
\author{Ernst Fischer\\Auf der Hoehe 82, D-52223 Stolberg, Germany}
\date{e-mail e.fischer.stolberg@t-online.de}
\maketitle

\begin{abstract}
Condensation in the outer regions of decaying supernovae is an efficient
source of dust with grain size up to more than $1 \rm{\mu m}$. The largest
grains leave the parent galaxy, thus forming "grey" intergalactic dust, which
can explain the observed dimming of light from distant supernovae without
invoking cosmic acceleration.\\

\end{abstract}
\section {Introduction}
During the last years two groups have investigated the behavior of distant
supernovae with the result that their luminosity is lower than expected by
standard cosmological models. In a number of papers (Perlmutter et
al.\cite{perl2}), (Schmidt et al.\cite{schmidt}; Riess et al.\cite{riess1})
both groups ascribe this luminosity lag to the action of a cosmological
constant, that means to the existence of a medium with repelling gravity,
which leads to an accelerated expansion of the universe. The authors argue
that neither evolution effects nor dimming by intervening matter can be
responsible for the weakening of the energy flux.

The main argument against extinction by cosmic dust, stressed by the two
groups, is that dust would lead to an additional reddening of the spectra,
which is not observed. In the reddening argument it is always assumed that
the intervening dust grains have approximately the same size distribution as
in our galaxy, thus producing the same strong wavelength dependence of
extinction.

There has been some criticism on this argument by Aguirre \cite{aguirre1},
who describes a method, by which dust consisting of small carbon needles can
be formed, showing less reddening than dust of galactic type. But as Riess et
al.\cite{riess2} have shown in a recent paper, even the reddening by the
carbon needles proposed by Aguirre (size $0.1\rm{\mu m}$) would be too strong
to be consistent with observations. In a more recent paper Aguirre
\cite{aguirre2} argues that dust particles formed in galaxies can be ejected
by radiation pressure. As this process favors the ejection of larger
particles and disfavors their corrosion, the intergalactic dust produced in
this way is less reddening than its galactic source. He shows that cutting
the galactic grain size distribution at $>0.1\rm{\mu m}$ results in reddening
compatible with observations.

For dust to be really "grey" in the spectral range observed in the distant
supernova search projects, the particle size should be in the order of
$1\rm{\mu m}$ . To explain the observed extinction of 30\% at z=0.46 (Riess
et al.\cite{riess2}) this would require that a mass corresponding to about
0.2\% of the critical density is contained in such dust grains. This can
easily be estimated, assuming that the dust consists of spherical particles
of similar composition as those in our galaxy (silicates) with a specific
mass around $2\rm{g/cm}^3$. If the density of ordinary matter in the universe
is close to 0.2 of the critical density, this value corresponds to a
"metallicity" close to that observed in the solar system.

Though the existence of such grey dust cannot be ruled out by any observed
effect, it has not yet been verified by other observations as well. By some
straight forward considerations we will propose a mechanism in this paper, by
which intergalactic dust can be formed, and which explains the characteristic
difference in particle size compared to galactic dust. One even had to look
for some new physics, if the extinction by grey dust should not exist. In
contrast to the paper of Aguirre \cite{aguirre2} in the mechanism proposed
here the dust grains directly condense in an environment flowing at
velocities exceeding the escape velocity at the site of formation.

\section {Production of cosmic dust}
It is generally believed that the chemical elements, from which cosmic dust
is composed, are produced predominantly by explosions of stars, which have
exhausted their nuclear fuel of hydrogen and helium. To understand the
process of dust formation, we must look somewhat closer at the processes
occurring during the collapse and subsequent explosion, observed as a
supernova. While in type Ia supernovae (SNIa) it is assumed that a white
dwarf exceeds its stability limit by accretion of mass and is completely
destroyed in the explosion, other types of supernovae occur, when high mass
stars collapse into a neutron star, expelling their outer shell in an
explosive event. Common to both explosions is, that matter in the order of
one solar mass consisting of elements heavier than He is ejected with
velocities up to 10000 km/sec.

As has been verified by several observations, SNIa form a group with well
established properties. If our understanding of SNIa is correct by principle,
the mass involved is close to the Chandrasekar limit and is completely
ejected in the explosion. The absence of hydrogen lines in the spectrum
indicates that the first steps of nuclear fusion are completed. Also a large
fraction of C, N and O should have been transformed into heavier elements. To
simplify the model of dust formation we restrict our considerations to the
processes in SNIa, to show that these processes can explain, why grey
intergalactic dust should exist in a quantity as mentioned above.

We will consider the effects following the explosion of a "template"
supernova as described by Perlmutter \cite{perl1}, that means an SNIa with a
peak magnitude $M=\rm{-19.5}$ and a decay time $dM/dt$ of 0.06m per day. The
expanding gas cloud is assumed to have a mass close to the Chandrasekar limit
( $M_{\rm {C}}=2.8 \times 10^{33} \rm g$), the element abundances being those
of the solar system, but without the presence of H and He. The radial
expansion velocity is assumed constant in the order of $u$ = 5000 km/sec.
Only in the very first moments of the explosion energy is produced by fusion
reactions and lost by emission of neutrinos, while during the period of decay
radiation from the outer layers of the expanding cloud is the only loss
mechanism. When the radius of the cloud is known, as long as the plasma is
optically thick, the surface temperature can immediately be estimated from
the measured luminosity by the relation
\begin{equation}
L=4 \pi R^2 \sigma T^4=10^{\frac{M_{\rm{peak}}-M}{2.5}} \times L_ {\rm{peak}}
\end{equation}
($M$ is the absolute bolometric magnitude; $ \sigma $ is the
Stephan-Boltzmann constant).

Using the relation $R=u \cdot t$, for the data given above recombination into
neutral atoms starts (at T=10000K) in the surface layer at about 20 days
after maximum brightness and is nearly complete after 50 days (4000K). Though
at this time the visible magnitude begins to deviate from the linear slope,
as the outer layers of the plasma become optically thin, we can assume that
the linear decrease of the bolometric magnitude continues for a longer period
of time, as the radiation loss is shifted more and more to IR.

Using this extrapolation we can calculate the time dependence of the
thermodynamic equilibrium composition of a gas of solar element abundances
(without H and He; C, N and O abundances reduced to 50\%). Assuming constant
pressure in the expanding gas sphere, the element concentrations are
determined by the total mass and the actual sphere radius. The results show
that molecules begin to form after 75 days (SiO and CO at 2500K). 105 days
from intensity maximum the temperature has dropped to 1500K. At this
temperature the first condensed phases ($\rm{MgSiO}_3$ and $\rm{SiO}_2$)
would occur under LTE conditions. ($\rm{Fe}_2\rm{SiO}_4$ would solidify at
1100K).

At that time the radius of the cloud is about $4.5\times 10^{15}\rm{ cm}$,
the particle density in the order of $10^9 \rm{cm}^{-3}$. Under these
conditions homogeneous nucleation cannot be expected in the hot gas, but we
must keep in mind that the cloud does not expand into vacuum, but into the
interstellar medium (ISM), which already contains dust grains, which are
swept up by the expanding gas and can serve as a nucleation seed.

Growth of the dust grains will preferably take place in the time period
immediately following the onset of supersaturation, as the number of
molecular collisions decreases with the third power of the cloud radius and
in addition with the square root of the gas temperature. To estimate the
final size, which a dust grain of initial radius $r_0$ can reach, we have to
calculate the total number of collisions between the grain and the
surrounding molecules. For simplicity we assume that there is only one type
of molecules with molecular mass $m$ and that the particle density $n$ is
constant in the gas cloud: $n=M_C/(4/3 \pi R^3) / m$. Assuming hard core
collision cross section, the mass of a spherical particle of density $
\rho_{\rm{D}}$ changes at a rate
\begin{equation}
\frac {d}{dt} \left( \frac{4}{3} \pi r^3 \rho_{\rm{D}} \right) = \pi r^2 n m v_{\rm{th}}.
\label {growth}
\end{equation}
$v_{\rm{th}}=\sqrt{3kT/m}$ is the mean thermal velocity (the sticking
coefficient has been set to 1). The time dependence of the temperature can be
approximated by a power law $T/T_{\rm{0}}=(r/r_{\rm{0}})^{\rm{-\alpha}}$ with
$\rm{\alpha}$ being close to one for the set of parameters used in the
example. Using the relation $R=u \cdot t$, eq.\ref{growth} can be integrated,
starting from the time of beginning supersaturation to infinity with the
result
\begin{equation}
r_{\infty}=r_{\rm 0}+\frac{3}{40\rm{\pi}}\frac
{M_{\rm{C}}v_{\rm{th_0}}}{\rho_{\rm{D}}u R_{\rm0}^2},
\end{equation}
The index 0 refers to the values at the time of beginning nucleation.
Inserting the data of the example leads to a final radius of about
$2.8\rm{\mu m}$. This value should be regarded only as an order of magnitude
estimation, as the derivation contains very crude simplifications. It is
assumed that all the material contained in the gas cloud is transformed into
species, which can condense into solid particles near 1500K. Solid species
occurring at 800K would reach only half the size, as the cloud has expanded
further in the mean time. The second assumption, which has been made, is that
the density of molecules in the cloud is not reduced by the condensation
process itself.

Whether this condition is satisfied, depends on the number of particles,
which form in the cloud. The order of magnitude should be equal to the number
of seed grains swept up from the ISM. The density of the ISM and its dust
content may vary strongly. Taking the orders of magnitude near the solar
system as a guide line, the density is in the order of $10^{-23}\rm{g/cm}^3$
with a dust content in the order of 1\%. Assuming a mean particle size of
$0.01\rm{\mu m}$ this leads to a grain number of $10^{-7}\rm{cm}^{-3}$. Most
of the grain growth in the expanding cloud will be completed, before the
cloud has reached a volume of $10^{48}\rm{cm}^3$. This means that the number
of grains would be in the order of $10^{41}$. If all the material contained
in the SN would condense onto these grains, they would grow to a mass of
about $3\times 10^{-8}\rm{g}$ corresponding to a radius of $15 \rm{\mu m}$.
Thus we can conclude, that it is the collision rate and not the availability
of material, which determines the final size of the dust particles. Grains,
which start growing at the beginning of supersaturation can easily reach a
size of more than $1 \rm{\mu m}$. Seed grains swept up at later times will
reach correspondingly smaller size.
\section {Size selection}
At the time of formation the radial velocity of the grains equals the
expansion velocity of the gas cloud. If they were ejected into empty space,
all grains would finally leave their parent galaxy, unless they are emitted
in the direction towards the center of the galaxy. For an SNIa exploding at
solar distance (7.7 kpc) from the center in an arm of a spiral galaxy of say
$10^{45} \rm g$, the escape velocity is about 500 km/sec, much less then the
velocity of the dust grains.

But in real galaxies the grains have first to traverse the ISM, before they
reach free space. In every collision with the gas atoms they lose momentum.
Even if the direction of emission is perpendicular to the disk plain, the
column density of the gas will be in the order of $n_{\rm{c}}=10^{21}
\rm{cm}^{-2}$.

In every collision with a hydrogen atom the momentum loss of a grain is
$-m_{\rm H}\cdot u$, as the velocity of the gas atoms is small compared to
the grain velocity. The momentum loss rate of a spherical grain of radius $r$
and density $\rho_{\rm D}$ thus is
\begin{equation}
\frac{4 \pi}{3} \rho_{\rm D}r^3\frac{du}{ds}=-n\cdot m_{\rm H}\cdot \pi r^2 u,
\end{equation}
or integrating along the path of the grain:
\begin{equation}
u_{\rm{\infty}}=u_{\rm0}\cdot \rm{exp} [-3m_{\rm H}n_{\rm c} /(4 \rho_{\rm
D}r)].
\end{equation}
If, as estimated above, the initial velocity is ten times the escape
velocity, for a grain to leave the galaxy the exponential factor has to be $>
0.1$. For a column density of $10^{21}\rm{cm}^{-2}$ this means $r >
2.5\rm{\mu m}$. Though this value can only be regarded as an order of
magnitude estimation, it clearly shows that the ISM acts as an effective
filter, which allows only the largest grains to escape, while the smaller
particles are thermalized to form the seed for condensation in subsequent
generations of SNe. The limit is just in the range, which has to be expected,
when the dust grains escaping from the galaxy show "grey" extinction, while
the remaining dust is highly selective.

Aguirre and Haiman \cite{aguirre3} have argued that too large dust grains
will absorb too much of the visible and UV radiation from stars, which is
then re-emitted in the far IR. This emission might be in conflict with
observations. For graphite particles they estimate an upper size limit near
$1\rm{\mu m}$. Due to the lower absorptivity of silicate grains, which are
expected to form in supernova remnants, such grains will heat up much less,
so that even with grains up to $r=2.5\rm{\mu m}$ the limit set by FIR
emission is not exceeded.
\section {Global distribution of dust}
The low scatter in the luminosity data of SNIa requires that any dust causing
the dimming must be uniformly distributed in space. Grains, which are formed
in gas clouds moving with initial velocities in the order of 10000 km/sec,
will reach free space retaining a considerable fraction of this velocity.
Thus within the age of the universe they can traverse even the largest voids
in space, so that a nearly uniform extinction is expected. This contrasts to
the mechanism of acceleration by radiation pressure, proposed by
Aguirre\cite{aguirre2}, where the final velocities are ten times lower.

By now it has been shown that the formation of large dust grains in SN
remnants is possible and that their escape from the parent galaxy is highly
selective with respect to grain size, but we have no good estimates, how
effective this mechanism of dust formation is, how many stars undergo an SN
explosion, and still more, what was the SN rate, when the galaxies were
young.

Additional information can be obtained, however, from the x-ray spectra of
galaxy clusters. In galaxy clusters dust grains, which leave their parent
galaxies, are still confined in the gravitation potential of the cluster. A
rich cluster may contain more than 1000 galaxies. But the mean distance of
the galaxies from the cluster center is only 100 times larger than the
distance of the SN, in which the dust is produced, from the center of the
individual galaxy. Thus ten times higher momentum is required to leave the
entire cluster, with the consequence that most of the dust is captured in the
intracluster medium (ICM).

As is well known from x-ray observations (see e.g. Sarazin \cite{sarazin2}),
hot gas with particle densities of $10^{-2}$ to $10^{-3}\rm{cm}^{-3}$ and
temperatures between $2 \times 10^7$ and $10^8 \rm K$ fills the space between
galaxies in clusters. By collisions with the high energy gas atoms the dust
grains will be eroded and finally resolved in the gas. An order of magnitude
estimation of the lifetime of grains can easily be obtained (similar to the
growth rate in eq.\ref{growth}) from the collision rate and the efficiency of
the erosion process. Detailed calculations of the sputter yield of various
atoms colliding with silicate grains have been carried out by P.W.May et al.
\cite{may}. Typical yields in the energy range considered here are in the
order of $\eta =10^{-2}$. For a spherical grain of radius $r$ the rate of
mass loss is
\begin{equation}
\frac{d}{dt}\left( \frac{4}{3} \pi r^3 \rho_{\rm D} \right)=-\eta m_{\rm a} n\cdot \pi r^2 v_{\rm{th}},
\label{eros}
\end{equation}
where $m_{\rm a}$ is the mass of the eroded atom and $v_{\rm{th}}$ is the
thermal velocity in the gas, which is of the order $10^8 \rm{cm/sec}$. From
eq.\ref{eros} we find by integration $\tau =r_0 \rho_{\rm D}/(\eta m_{\rm
a}nv_{\rm{th}})$ for the lifetime of a grain with initial radius $r_0$.
Typical data as discussed above result in lifetimes in the order of $10^{15}$
sec for $1\rm{\mu m}$ grains. That means that a large fraction of the grains
is completely resolved, so that from the metallicity of the cluster gas we
can immediately derive the amount of dust, which has been ejected during the
existence of the galaxy cluster.

X-ray spectra of most clusters show metallicities around half that of the
solar system (Sarazin \cite{sarazin1}). Thus, if emission of dust from
galaxies is the dominant mechanism for enrichment of heavy elements in the
intergalactic medium, we can conclude, that the overall abundance of these
elements in free space today should be of the same order as in the ICM. But
in free space, where no erosion takes place, the heavy elements remain bound
in large dust grains.

Of course there is the possibility that also outside of galaxies and clusters
space is not empty, but filled with hot gas, which might for instance be
ejected by the jets of quasars. But even if such an intergalactic plasma
exists with a density close to the critical closing density of the universe,
it would be 1000 times more dilute than the intracluster gas, so that the
erosion time of dust grains would be in the order of $10^{18}$ sec,
comparable to the age of the universe. Thus most grains, that have escaped
into free space since the formation of galaxies, will still be there, causing
dimming of the light from distant galaxies just at the level observed in the
light curves of SNIa.

The study of distant supernovae has not proved the existence of a by now
undetected medium with repulsive gravity. However, for the first time it has
been proved that the space between galaxies is not as empty as thought
before, but that it contains matter, which can considerably influence the
observations of distant regions in space.
\\

\end{document}